\documentclass[3p,12pt]{elsarticle}

\usepackage{amssymb}
\usepackage{amsmath}
\usepackage{graphicx}
\usepackage{esint}
\usepackage{color}

\usepackage{array}
\usepackage{multirow}

\PassOptionsToPackage{hyphens}{url}
\usepackage[pdfencoding=auto,psdextra,colorlinks=true,pdfauthor=author]{hyperref}
\usepackage{bookmark}
\biboptions{numbers,sort&compress}

\journal{Technology in Society}

\date{May 31, 2023}

\begin{document}

\begin{frontmatter}



\title{Quantitative dynamics of design thinking and creativity perspectives in company context}

\author[label2]{Georgi V. Georgiev}
\ead{georgi.georgiev@oulu.fi}
\affiliation[label2]{organization={Center for Ubiquitous Computing, Faculty of Information Technology and Electrical Engineering, University of Oulu},%Department and Organization
            %addressline={Linnanmaa campus, Erkki Koiso-Kanttilan katu 3}, 
            city={Oulu},
            postcode={FIN-90014}, 
            %state={},
            country={Finland}}

\author[label1]{Danko D. Georgiev}
\ead{danko.georgiev@mail.bg}
\affiliation[label1]{organization={Institute for Advanced Study},%Department and Organization
            addressline={30 Vasilaki Papadopulu Str.}, 
            city={Varna},
            postcode={9010}, 
            %state={},
            country={Bulgaria}}

\begin{abstract}
This study is intended to provide in-depth insights into how design thinking and creativity issues are understood and possibly evolve in the course of design discussions in a company context. For that purpose, we use the seminar transcripts of the Design Thinking Research Symposium 12 (DTRS12) dataset ``Tech-centred Design Thinking: Perspectives from a Rising Asia,'' which are primarily concerned with how Korean companies implement design thinking and what role designers currently play. We employed a novel method of information processing based on constructed dynamic semantic networks to investigate the seminar discussions according to company representatives and company size. We compared the quantitative dynamics in two seminars: the first involved managerial representatives of four companies, and the second involved specialized designers and management of a design center of single company. On the basis of dynamic semantic networks, we quantified the changes in four semantic measures---abstraction, polysemy, information content, and pairwise word similarity---in chronologically reconstructed individual design-thinking processes. Statistical analyses show that design thinking in the seminar with four companies, exhibits significant differences in the dynamics of abstraction, polysemy, and information content, compared to the seminar with the design center of single company. Both the decrease in polysemy and abstraction and the increase in information content in the individual design-thinking processes in the seminar with four companies indicate that design managers are focused on more concrete design issues, with more information and less ambiguous content to the final design product. By contrast, specialized designers manifest more abstract thinking and appear to exhibit a slightly higher level of divergence in their design processes. The results suggest that design thinking and creativity issues are articulated differently depending on designer roles and the company size.
\end{abstract}


\begin{highlights}
\item Design conversations are analyzed with an information-processing method based on semantic networks.

\item Design thinking in Korean companies is understood based on temporal dynamics of semantic networks.

\item Dynamic changes in abstraction, polysemy, information content and semantic similarity are quantified.

\item Individual design thinking is articulated differently depending on the designer role and company size.

\item AI co-creative partner programs employing semantic networks could assist the work of human designers.
\end{highlights}


%Published in: Technology in Society 2023; 74: 102292. DOI:~\href{http://doi.org/10.1016/j.techsoc.2023.102292}{10.1016/j.techsoc.2023.102292}

\end{frontmatter}


%1
\section{Introduction}

%1.1
\subsection{Semantic analysis of design thinking}

The cognitive processes that occur inside designer's mind and underlie the extraordinary ability of solving creative problems are referred to as \emph{design thinking} \citep{Cross2023,Nagai2003,Nagai2012,Georgiev2012,Razzouk2012}.
Language is a powerful and compelling data source for analyzing knowledge \citep{Sarica2023} and mental processes \citep{GoldmanEisler1958,Ericsson1980,Taura2012}, including design thinking \citep{Cross2023,Adams2015,Georgiev2018}, conversations about design \citep{Kan2017}, and evaluations of creative projects \citep{Li2022}. Computational processing of language provides critical insights into design-thinking and problem-solving processes \citep{Dong2009}. These processes can be modeled by analyzing semantic networks that offer a structured representation of human thinking as an associative system of concepts and semantic connections \citep{Hartley1997,Boden2004,Baronchelli2013}. Nodes in a semantic network represent specific concepts or semantic entities such as words, whereas links refer to mental (semantic) relations, exemplifying how concepts can be accessed from one another \citep{Boden2004,Han2021}. Dedicated semantic networks constructed with information extracted from textual data, can be functionally and structurally focused, providing a network representation of external knowledge \citep{Liu2022}. Semantic features and patterns have proven their utility in modeling design descriptions \citep{Sarica2023}. Knowledge connections and construction in networks can promote design thinking and information processing \citep{Shen2021}.

Semantic analysis based on lexical chains has been used to capture competing interests, their reconciliation, and the resulting agreement in design problem solving \citep{Dong2007,Dong2009}. Forms of language for expressing judgments and identifying semantic resources in linguistic appraisals in design are detailed by \cite{Dong2009}. Moreover, semantics-based approaches to design analysis \citep{Dong2005,Dong2007} emphasize that word relations are essential to understanding design thinking. Noun phrases evolve over the course of a design project and can serve as useful surrogates for measuring the early phase of the mechanical design process when multiple alternatives are generated \citep{Mabogunje1997}. The formation of different noun--noun combinations and noun--noun relations reflects the underlying conceptual combination of design ideas.
Semantic networks of nouns constructed from verbal data can also be used to investigate creativity in conceptual design and to simulate difficult-to-observe design-thinking processes \citep{Taura2012}. Semantic networks are employed as knowledge bases for purposes such as knowledge retrieval, concept association, and the expansion or execution of queries \citep{Han2021}.

Alternative qualitative methods for investigating design thinking, such as linkography \citep{Goldschmidt1990,Goldschmidt2014,Goldschmidt2016,Kan2017} or thematic analysis \citep{Braun2022}, rely on qualified design experts (human raters) to recognize the decisions, activities, and themes that occur during the design-thinking process. Typically, the interrater agreement (consensus) and interrater reliability (consistency) deviate significantly from unity \citep{Fleenor1996} and vary with the academic background or years of research experience of the human raters \citep{Belur2018}, which introduces a subjective error into the analyzed data and limits the reproducibility of results by independent research teams \citep{Alavi2022}. In addition, qualitative data are comprehensible to humans, but are unusable by computer algorithms, which can only interpret and work with accurate quantitative data.

In this study, we employ semantic networks as a robust, objective, and highly reproducible alternative to qualitative methods, because the quantification of different semantic measures can be fully automated and executed by a computer program in the absence of any design experts or human raters. The computational method employs Natural Language Processing scripts to extract nouns from conversation transcripts, construct dynamic semantic networks of nouns, and automatically quantify different semantic measures of interest based on WordNet~3.1 (https://wordnet.princeton.edu/). By eliminating human raters from coding the design conversation transcripts, we can ensure maximal 100\% reproducibility in the constructed dynamic semantic networks and their corresponding quantitative semantic measures.

%1.2
\subsection{Measures on semantic networks for analysis of design thinking}

Recent research has fostered a number of approaches to analyze design thinking based on semantic networks \citep{Georgiev2010,Georgiev2018,Han2020} or information use in the thinking process of designers based on the dynamics of networks of linked data \citep{Cash2014}.
There are three main advantages to using dynamic semantic networks to model human creativity.
First, the method is applicable to studying any cognitive processes occurring in the human mind that can be verbalized \citep{Ericsson1980}.
Second, a large number of information theoretic semantic measures can be computed from transcribed design conversations using available natural language processing Python scripts \citep{Bird2009} and WordNet \citep{Georgiev2018}.
Third, dynamic semantic networks can be computer-generated, visualized, and used by designers in real time.

From a structured view of information processing, semantic networks or knowledge graphs can be explicit representations of personal knowledge about the world \citep{Sarica2023}.
Design knowledge representation has informative and reasoning advantages when networks with semantic relationships of design terms are used, and semantic measures on such networks have been used for design process characterization \citep{Siddharth2021}.

Previous study analyzed the Design Thinking Research Symposium~10 (DTRS10) dataset of design problem-solving conversations, with a focus on the temporal dynamics of semantic factors that quantify real-world human problem-solving processes in the design educational context \citep{Georgiev2018}.
Using a large set of semantic measures, that study found that the dynamics of three semantic factors---semantic similarity, information content and polysemy---can \emph{predict the success} of generated ideas in an educational context \citep{Georgiev2018}.
In particular, the divergence of semantic similarity, increased information content, and decreased polysemy were significant features of successful design solutions.

%1.3
\subsection{Information exchange when designing and analyzing designing}

Information exchange can be achieved in multiple ways, including through mechanical, electrical, and natural interactions \citep{Zhu2022}. The most common methods of natural interaction are gesture, eye-movement, and voice interactions \citep{Becker2020}. Here, we focused mainly on the information exchange through language \citep{Ericsson1980} that occurs between participants of design seminars in a company context \citep{Christiaans2018}.

Semantic analysis of technology-supported idea generation of professional designers shows that semantic factors precisely delineate the differences between design thinking sessions supported by Information and Communications Technology (ICT) and non-ICT-supported sessions \citep{Becattini2020}.
Semantic factors characterizing discussions when designing can account for specific creativity aspects such as originality or feasibility of the outcome \citep{Casakin2021}.
Therefore, information exchange in the form of conversations or discussions in design thinking is essential for understanding, aiding, and improving the design thinking as a process \citep{Goldschmidt2014,Han2021}.

When analyzing or discussing design thinking, information exchange is essential for aiding and improving design thinking in practice.
For example, information-seeking, analysis, and reflection practices positively influence team performance, and most importantly, outcome innovation \citep{Wang2020}.
\cite{Lee2018} also compared conversations describing designing and design sessions about design thinking.
Their analysis focused on categorizing issues, and the findings showed that significantly more functional and structure-related issues were discussed when describing the designing process.
Discussions about design might provide particular insights into the overall mechanisms of the design-thinking process.
In a recent study, roles in the design-thinking process were distinguished in terms of communication and its cohesion \citep{Song2022}.
Integration of different stakeholder perspectives is particularly relevant for the case of small and medium-sized enterprises (SMEs) \citep{Selamat2021}.

%1.4
\subsection{Research questions}

This study is intended to provide in-depth insights into how design thinking and creativity issues are understood and evolve in the course of design discussions between professionals, intra- and inter-organizationally, based on semantic networks and their dynamics.
Seminar transcripts from the Design Thinking Research Symposium~12 (DTRS12) dataset were used.
The analysis focuses on the differences in the networks and graphs pertaining to participants from big companies and SMEs.
These differences and their dynamics over time would shed light on the reasons for the minority role of design and creativity in the latter companies from the South Korean perspective, as outlined in the Introduction of DTRS12 \citep{Christiaans2018}.
Furthermore, they offer new avenues for intervention to address these and related issues.

Our main research questions are as follows: (1) ``What are the temporal dynamics of quantitative semantic measures, such as abstraction \citep{Nagai2009}, polysemy \citep{Georgiev2014}, information content \citep{Sanchez2012}, and semantic similarity \citep{Resnik1999}, in the individual design-thinking process of designers in the company context?,'' (2) ``How do the observed dynamics of those semantic measures compare to the previously reported \citep{Georgiev2018} association between divergent thinking and creativity in the educational academic context?'', and (3) ``How are design thinking and creativity issues articulated in companies of different size and by different designer roles?''

The practical significance of having a quantitative description of design creativity in the form of dynamic semantic networks is that they can be reverse engineered for the future development of artificial intelligence (AI) co-creative partner systems \citep{Kim2023}. In contrast to qualitative data, which are only comprehensible to humans \citep{Braun2022,Dym2005}, quantitative data can be used by computer programs for control and navigation toward a certain computational goal. For example, the general qualitative statement that creativity is exercised through divergent combination of distant ideas \citep{Guilford1957} is poorly defined and does not exclude combination of ideas that are completely unrelated to the design problem that has been posed. In contrast, having a precise percentage of divergence that occurs in the work of human designers could be used by AI co-creative partners (computer programs) to evolve an initial sketch of a solution into a number of possible design solutions from which the human designer is able to choose. Because the adoption of AI co-creative partner systems will be spearheaded by commercial companies, before AI systems are available for personal use, the quantitative characteristics of design creativity in the company context are worth investigating, as they will lay the foundation for such AI development.

%2
\section{Selected semantic measures and their roles in the design process}

%2.1
\subsection{Abstraction in design thinking and problem solving}

Expert design thinkers engage in stepping back or moving between levels of abstraction \citep{Kokotovich2016}.
A low level of abstraction was observed when the design thinking is concrete
and directly related to the physical variables of the design. By contrast, a high level of abstraction is related to the essence of the design problem \citep{Kokotovich2016}. Abstraction has been observed to increase over lifetime and is related to the expertise levels of design thinkers. The
abstraction is inhibited in case of group design thinking \citep{Dong2013}.

Overall, compared to specific ways of thinking, abstract thinking can lead to open-ended and, therefore, novel ideas \citep{Ward2004,Saitta2013}.
Generalization and abstraction have been linked to the formation of semantic networks \citep{Khalil2022}. Furthermore, the generation of multiple unique ideas is enhanced by the availability of abstract text stimuli \citep{Goncalves2012}.
Therefore, abstraction is an essential feature of creative idea generation
and design thinking \citep{Welling2007}.

%2.2
\subsection{Polysemy in design thinking and problem solving}

Polysemy, the potential of a word to possess multiple meanings \citep{Taura2012}, has been investigated by scholars interested in the early stages of design problem solving.
The average polysemy of design ideas was found to be significantly correlated with their originality ratings in a design task in which a new design idea had to be generated from two given base concepts \citep{Taura2012,Yamamoto2009}.
\cite{Georgiev2014} identified polysemy as a characteristic feature of successful ideas considered in the final solution in design conversations. Polysemy demonstrates the multiplicity of significations in a designed object \citep{Dabbeeru2011}.

Overall, different manifestations of creativity seem to operate through concepts that exhibit high levels of polysemy.
Polysemy has been identified as an essential manifestation of the flexibility, adaptability, and richness of the meaning potential of a language \citep{Fauconnier2003}.
Polysemy also provides flexible patterns for cognitive operations that support creativity at various levels \citep{Nerlich2003}.
Moreover, creative inspiration can originate in polysemy, allowing for the examination of the diverse meanings of related concepts \citep{Zhang2014}.

%2.3
\subsection{Information content in design thinking and problem solving}

Information content is a quantitative measure of the amount of information transmitted by a specific language unit in a certain context \citep{Georgiev2018}.
Ontology-based computation of information content has shown great potential for analyzing problem solving, because it is better correlated
with human judgments than corpora-based computation of information content \citep{Sanchez2011}.
Words (concepts) with higher information content are less likely to occur in general contexts \citep{Meymandpour2016}. The information content between links using Shannon's concept of entropy was used to measure design fixation \citep{Gero2011}. The latter study illustrates how the theory can be applied to locate and measure design fixation during particular segments of design-thinking sessions \citep{Gero2011}.
Furthermore, information content has been useful for detecting high creativity scores during design sessions \citep{Kan2017} and determining the usefulness of solutions generated in design thinking \citep{Sen2010}.

%2.4
\subsection{Semantic similarity in design thinking and problem solving}

Semantic similarity quantifies the strength of the semantic relationships between pairs of words. In the design-thinking process, semantic similarity was related to the novelty of a design produced by combining two initial concepts \citep{Nomaguchi2019}.
Extensive analysis of real-world design problem solving has demonstrated
that convergence and divergence in design thinking are faithfully captured by the dynamics of semantic similarity in the constructed semantic networks of nouns \citep{Georgiev2018}. In particular, successful ideas manifest decreased semantic similarity and increased information content over time, a combination that is considered a hallmark of divergent thinking \citep{Georgiev2018}.
Latent Semantic Analysis can also be used to assess semantic similarity on a macro level between different texts in the context of problem solving \citep{Quesada2002}. The role of semantic similarity in design is highlighted by the finding that the degree of similarity or dissimilarity of noun--noun combinations is related to creativity by yielding emergent properties in idea generation \citep{Wilkenfeld2001}.

Conceptual distances, which represent the degree of similarity between ideas or concepts, have been employed in a combinational creativity approach \citep{Han2020}. Similarities or conceptual distances can be evaluated using different approaches \citep{Han2021}. However, in general, the conceptual distances between the base and additional ideas relate to the degree of creativity of the idea combination, suggesting that distantly related ideas are potentially more creative \citep{Han2020}.

%3
\section{Methods}

%3.1
\subsection{Dataset overview and purpose of the design seminars}

This study uses the Design Thinking Research Symposium~12 (DTRS12) ``Tech-centred
Design Thinking: Perspectives from a Rising Asia'' dataset.
The starting point of the dataset is how Korean companies implement design thinking and what role designers will play now and in the future.
Moreover, this coincides with the current global need for academia to cooperate with industry in developing knowledge and skills that are more readily available and applicable \citep{Christiaans2018}.

\begin{figure}[t!]
\begin{centering}
\includegraphics[width=\textwidth]{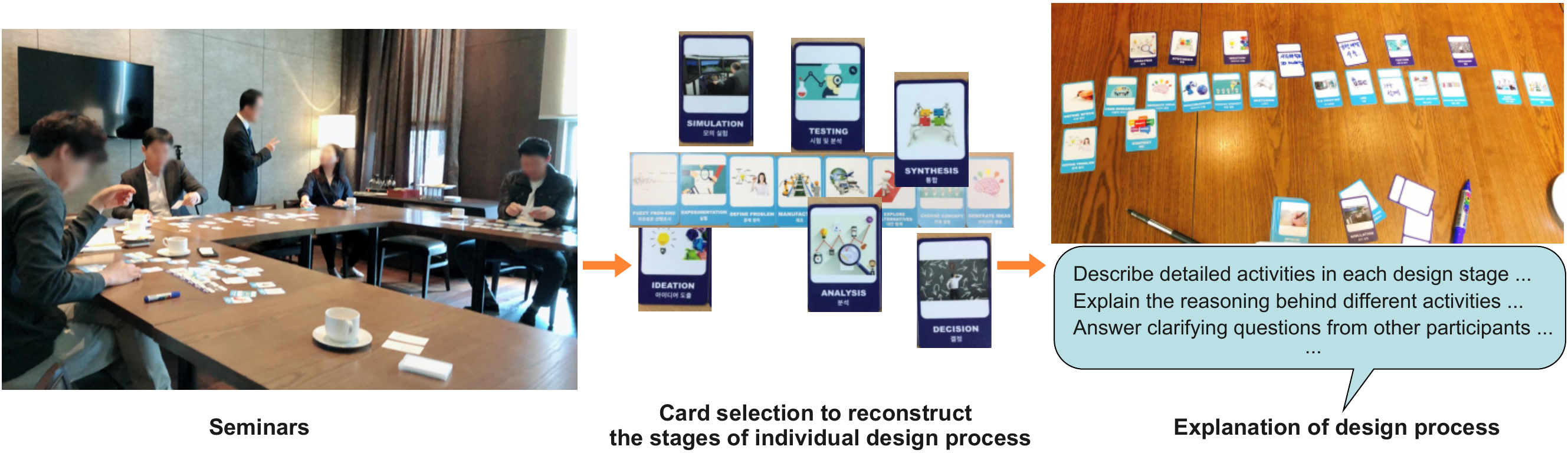}
\par\end{centering}
\caption{\label{fig:1}General overview of the design seminars in the DTRS12 dataset. Participants in the seminar used a large set of cards to reconstruct the stages of their individual design-thinking process, after which they described the various design activities and explained the rationale behind them.}
\end{figure}

The in situ data of the dataset comprise recorded transcripts of two seminars (workshops) with high-ranking managers and company employees. The workshops were organized with participants from five Korean companies, ranging from conglomerate (chaebol) to medium-sized companies \citep{Christiaans2018}.
The first seminar involved representatives from four companies, including a conglomerate, two smaller companies, and a marketing consultancy. Hereinafter, it is denoted as the ``seminar with four companies'' (Fig.~\ref{fig:1}).
The second seminar was conducted with a large Korean pharmaceutical company, which has a broader product range than only medical related products. Hereinafter, it is denoted as the ``seminar with design center.''
The original language used in the DTRS12 workshop was Korean. The conversation transcripts in the DTRS12 dataset were provided both in Korean and in parallel English translation.

\begin{figure}[t!]
\begin{centering}
\includegraphics[width=\textwidth]{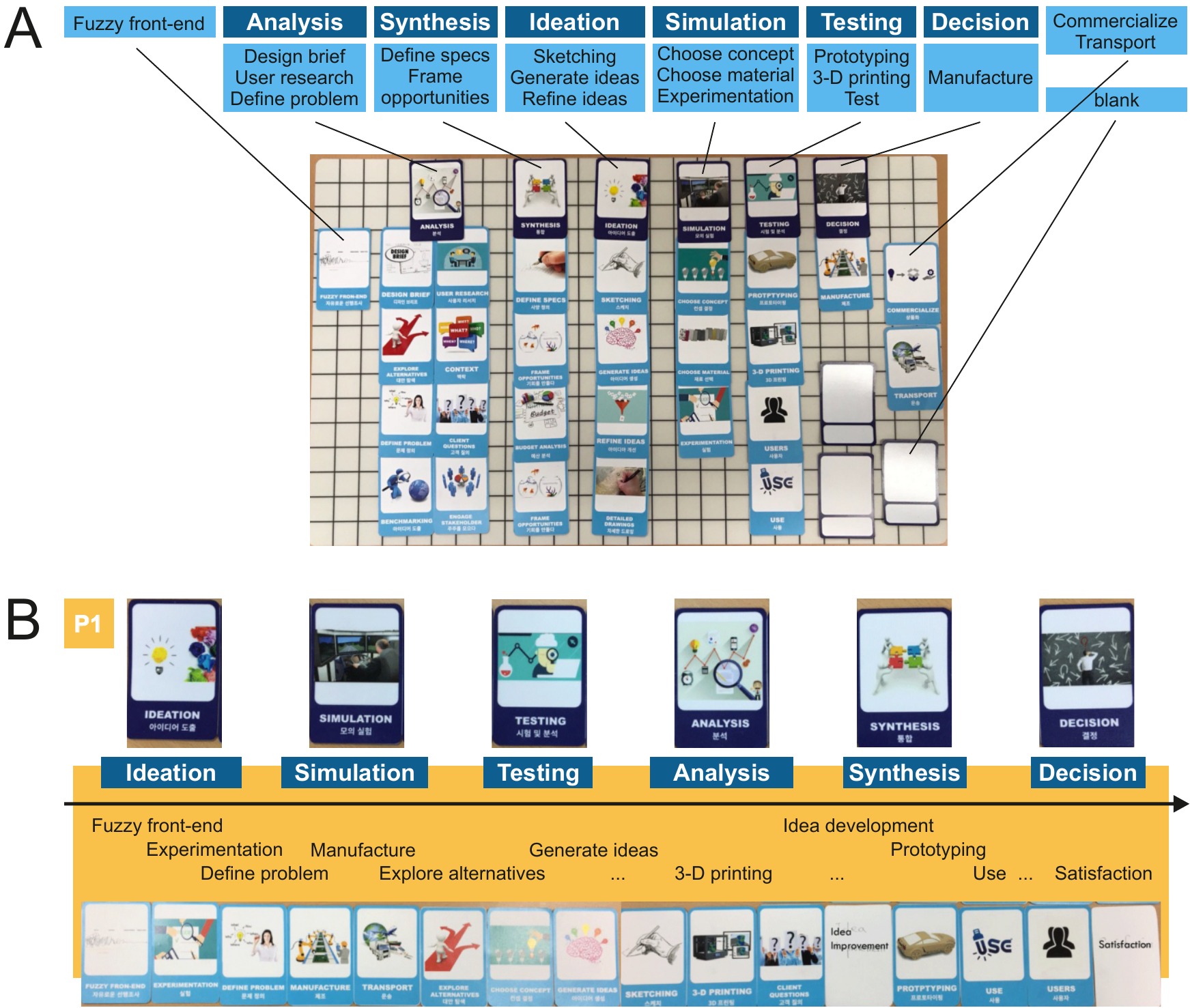}
\par\end{centering}
\caption{\label{fig:2}Large card set and an example of a reconstructed individual design-thinking process. (A) Dark blue cards in the large card set represent general stages of the design process, whereas light blue cards describe detailed activities. Empty (blank) cards were also included that could be used in case a design aspect or activity was missing. (B) The reconstructed individual design-thinking process of designer P1 consists of 6 stages ordered chronologically as: Ideation, Simulation, Testing, Analysis, Synthesis and Decision. Each design stage contains a number of corresponding detailed design activities.}
\end{figure}

The design seminars were intended for individual participants to reconstruct their design process into a chronologically ordered sequence of design stages and elaborate on their design thinking at each stage (Fig.~\ref{fig:2}). In both seminars, participants were aided by the same large card set, in which each card depicted a distinct activity or aspect of the design process. Each participant was asked to select a subset of cards from the large card set that best matched the stages of their own design process and explain their ongoing design thinking with the help of the given cards \citep{Christiaans2018}. The verbal reports obtained were both personal reflections on the design-thinking process and a chronological mental replay of the stages of the design process. This latter fact justifies our subsequent semantic analysis of the  DTRS12 dataset as a study of the design-thinking process in the company context, even though actual design conversations for the development of concrete commercial products were not provided in the DTRS12 dataset to protect company secrets.

\subsubsection{Seminar with four companies}

Participants with high-ranking managerial roles (\emph{design managers}) participated in the seminar with four companies (Fig.~\ref{fig:1}). These were from an SME (the participant is denoted hereinafter as N), from a large company (denoted as S), from a marketing consultancy (denoted as C), and from another SME (two participants denoted as H1 and H2). Two university researchers also participated in the seminars (denoted as R1 and R2), with minor facilitating roles in the conversations.

\subsubsection{Seminar with design center of a company}

In this seminar, there were five participants (\emph{specialized designers}) from a pharmaceutical company which has a broader product range than only medical related products \citep{Christiaans2018}. The roles of the specialized designers in the company were product design (hereinafter denoted as P1), graphic design (denoted as P2), intern (denoted as P3), creative design (denoted as P4), design center chief (not actively participating in the conversation, thus not further denoted here), and university researcher (denoted as R).

%3.2
\subsection{Semantic networks}

Semantic networks can be used to computationally model conceptual
structures and associations \citep{Hartley1997,Georgiev2010}.
The method employed in this study uses semantic networks that represent concepts (meanings and words) as nodes in a graph, and relationships as links between the nodes.
The understanding of an issue can be represented through a semantic network or graph. The process of discussing an issue and finding a solution can be understood using a semantic network that changes over time \citep{Georgiev2018,Casakin2021}.
A practical way to analyze discussions is by computing graph-theoretic measures from constructed graphs based on workshop discussion transcripts, in which the participants clarify design and creativity issues.

To construct dynamic semantic networks of nouns, we used the conversation transcripts from the DTRS12 seminars.
First, we cleaned the conversation transcripts by removing images, speaker names, and all indications of non-verbal expressions, such as laughter or exclamations.
Second, we processed the clean text using part-of-speech tagging performed by the Natural Language Toolkit (NLTK) \citep{Bird2009}
with the TextBlob library \citep{Loria2016}.
Third, we extracted singular and plural nouns in the order of their occurrence. Each noun was written onto a new line in plain .txt files.
Fourth, we processed all the nouns by converting the plural forms to singular forms using custom Python scripts, and excluded those nouns that were not listed in WordNet~3.1 \citep{Miller1995,Fellbaum1998}.\\

\begin{figure}[t!]
\begin{centering}
\includegraphics[width=\textwidth]{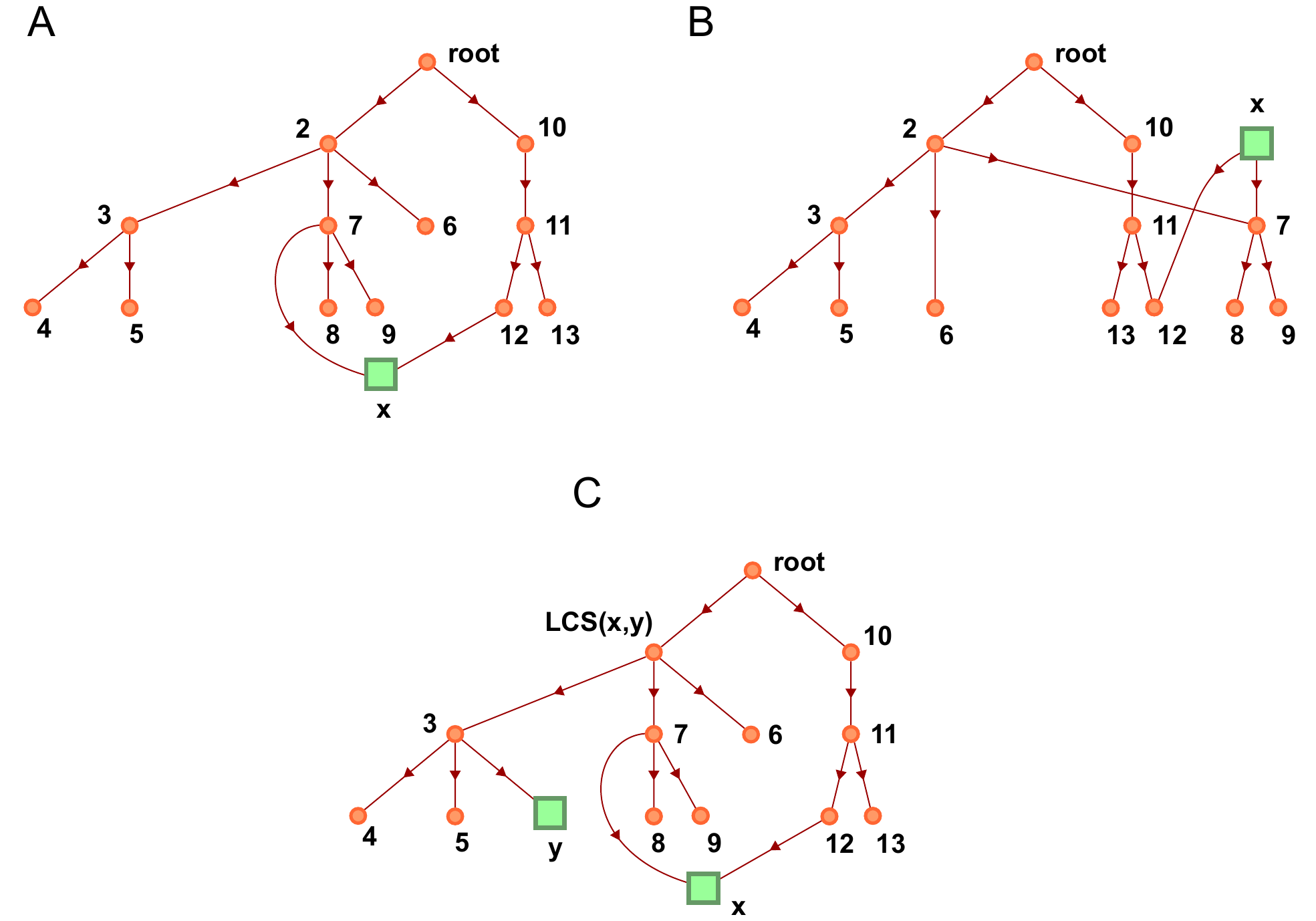}
\par\end{centering}

\caption{\label{fig:3}Semantic functions, which take word arguments in WordNet
3.1 fragment composed of meaning vertices (orange circles) and word
vertices (green squares). (A) Depth(x) = 3; \textbar{}Subsumers(x)\textbar{}
= 6; (B) Polysemy(x) = 2; \textbar{}Subvertices(x)\textbar{} = 4;
\textbar{}Leaves(x)\textbar{} = 3; Commonness(x) = 3/4; (C) LCS(x,y) is the lowest common
subsumer of $x$ and $y$; Depth{[}LCS(x,y){]} = 2.}
\end{figure}

The following is a sample of graph-theoretic functions and measures that are computed with the use of WordNet~3.1 hypernym--hyponym (is-a) hierarchy of nouns (Fig.~\ref{fig:3}),
which was rendered as a graph composed of \emph{word} nodes, \emph{meaning} nodes, and directed links between the nodes \citep{Georgiev2018}.

\begin{itemize}
\item[$\triangleright$] \emph{Path} is a finite sequence of links that connects a sequence of nodes \citep{Diestel2017}.

\item[$\triangleright$] \emph{Depth} is the number of meaning nodes located on the shortest path from the root meaning node to a meaning node that subsumes the word node of interest.

\item[$\triangleright$] \emph{Abstraction} is the complement to unity of the shortest path distance from the root to a meaning node subsuming the word of interest, normalized to the maximal shortest path from the root in the graph of meanings \citep{Georgiev2018}.
For example, in Fig.~\ref{fig:3}A the shortest path to word $x$ from the root is through nodes 2 and 7, with a distance of 2 links between the meaning nodes. The maximal shortest path from the root in the graph of meanings consists of 3 links between meaning nodes, whereas the maximum depth is 4 meaning nodes.

\item[$\triangleright$] \emph{Polysemy} is the number of direct links between a word node and its meaning nodes \citep{Georgiev2014}. For example, in Fig.~\ref{fig:3}B, the polysemy of the word $x$ is 2.

\item[$\triangleright$] \emph{Information content} is the bits of information carried by a node inside the graph. The normalized information content of a node is computed from the commonness of the node by the following formula $IC(x)=\log[\mathcal{C}(x)/\mathcal{C}_{\max}]/\log[\mathcal{C}_{\min}/\mathcal{C}_{\max}]$,
where $\mathcal{C}(x)$ is the commonness of $x$, $\mathcal{C}_{\min}$ is the minimal commonness and $\mathcal{C}_{\max}$ is the maximal commonness in WordNet~3.1 \citep{Sanchez2012}.

\item[$\triangleright$] \emph{Semantic similarity} is the information content of the lowest common subsumer (LCS) of two words \citep{Resnik1999}.
\end{itemize}

We are particularly interested in the quantitative description of human design thinking that can be inferred from these networks and graphs.
The temporal dynamics of semantic measures could identify real-world processes in human design thinking that are relevant to the evolution and outcomes of discussion and could provide insight into how communication affects the development of knowledge in the conversation, in particular, furnishing a structured representation of knowledge communicated during the conversation (explanation with cards activity).

%3.3
\subsection{Temporal dynamics of semantic measures}

\begin{figure}[t!]
\begin{centering}
\includegraphics[width=\textwidth]{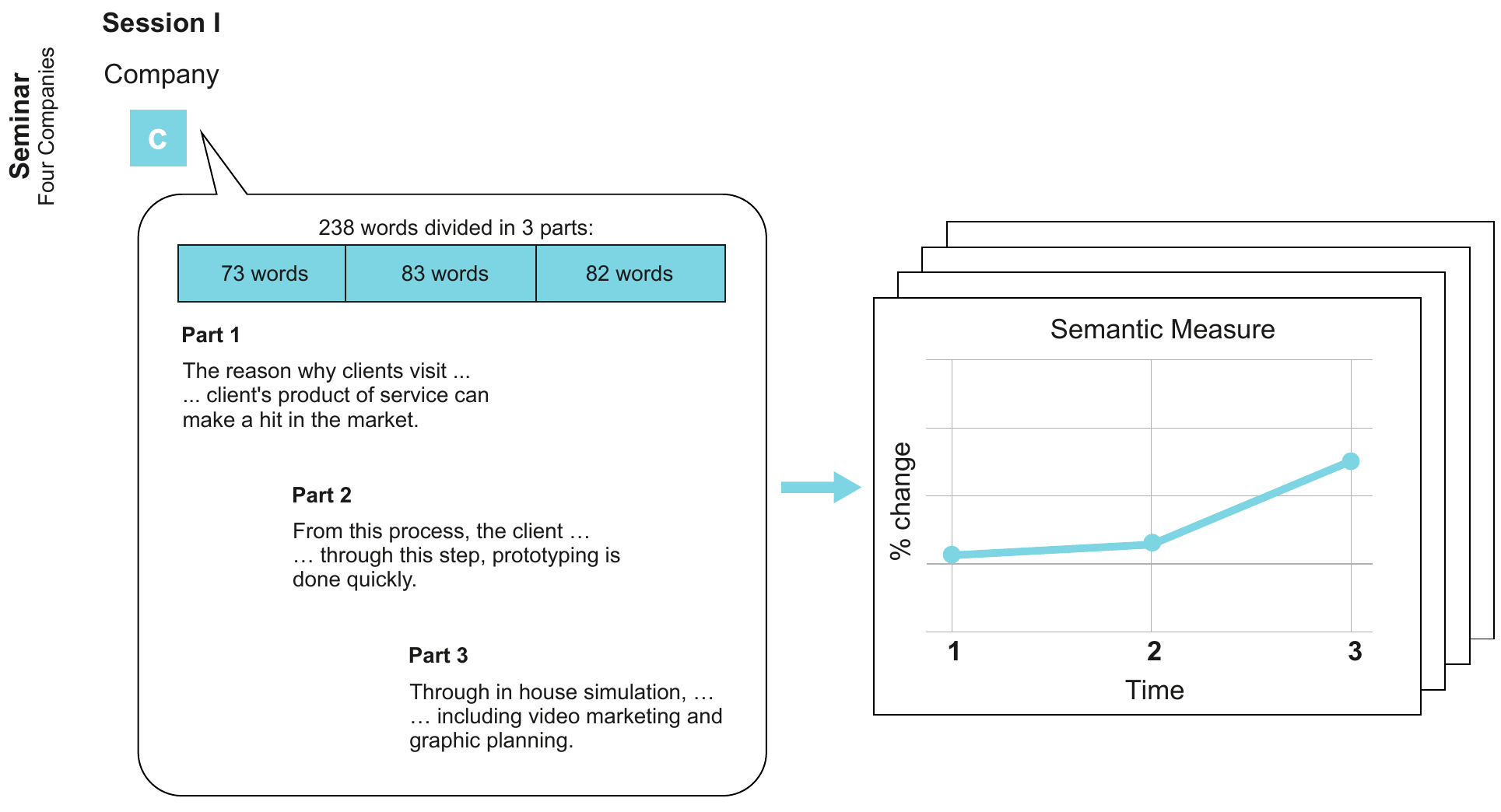}
\par\end{centering}

\caption{\label{fig:4}Example division of a conversation into three parts
to quantify dynamics.}
\end{figure}

To quantify the dynamics of understanding design and creativity issues, the conversations for each individual design-thinking process were divided into three approximately equal parts based on word count (Fig.~\ref{fig:4}).
This division into parts was made into whole sentences, ensuring that all conversation parts contained at least five distinct nouns.
The average values of the semantic measures were then calculated for each part, and three time periods were obtained $t\in\{1,2,3\}$.

We used the English translation of the DTRS12 dataset to construct the dynamic semantic networks of concepts. The quantitative results obtained from the English text are expected to be largely language-independent because of the general characteristics of the computational method employed, whose primary focus is on concepts regardless of the particular language into which these concepts are translated.
First, the concepts in the semantic network include only nouns, and not verbs, adjectives, or adverbs. Thus, the language grammar of sentences is irrelevant to the constructed semantic networks.
Second, while the temporal dynamics is sensitive to the order of appearance of the nouns, the design conversations were divided into temporal parts using whole sentences. Consequently, the method is not affected by the language-specific ordering of nouns in a sentence.
Third, the hypernym--hyponym (is-a) hierarchy of nouns in WordNet~3.1, is encoded in the core graph of meanings and their subordination. For example, the word ``sky'' with meaning vertex M09459612 is a kind of ``atmosphere'' with meaning vertex M09233511. The graph relation between the vertices  M09233511$\rightarrow$M09459612 is not affected by the translation of the monosemous words ``sky'' or ``atmosphere'' into another language. In fact, the word interface could be replaced by a pictorial one---for example, by replacing the word ``sky'' with a photo image of a blue sky---without affecting the essence of the constructed semantic network. The only possible difference between WordNets in different languages could occur for polysemous words, which have more than one meaning \citep{Georgiev2018}. Unfortunately, a direct assessment of the effect of polysemy on semantic networks constructed with the English WordNet~3.1 or the Korean WordNet (http://wordnet.kaist.ac.kr/) is precluded, because the Korean WordNet project is only in its initial state of development. For example, while the English WordNet subnet of nouns contains 82192 synsets (meaning vertices) and 158441 words (word vertices), the Korean WordNet currently contains only 9714 synsets and 8270 words. We believe that our demonstration of the utility of the English WordNet for the semantic analysis of English text will entice the development, completion, and successful use of WordNets in other languages.

%3.4
\subsection{Examples}

The examples listed in Table~\ref{tab:1} illustrate the possible temporal changes that can be quantified with dynamic semantic networks.
Information content is calculated according to a logarithmic formula using the commonness of a node (word) \citep{Sanchez2012} and semantic similarity is determined by the information content of the lowest common subsumer (meaning) of a pair of words \citep{Resnik1999}. This selection was grounded in previous research, in which these formulae demonstrated high statistical power and were significant predictors of the outcomes of design problem-solving conversations \citep{Georgiev2018}. Moreover, they are comparatively easy to calculate; thus, they can be easily implemented in the envisioned real-time analysis of design conversations.

The first column of Table~\ref{tab:1} lists a set of five initial words (nouns):
`business, design, process, purpose, view'. The four semantic measures
of this set, abstraction, polysemy, information content, and semantic
similarity, were calculated to be 0.767, 7.000, 0.498 and 0.521, respectively.
When in the course of a conversation, a new word is introduced, for example
`order', the overall values of the four semantic measures, abstraction, polysemy,
information content, and semantic similarity, become 0.769,
8.333, 0.477 and 0.447, respectively, resulting in increased abstraction
and polysemy, and decreased information content and semantic similarity.

\begin{table}
\caption{\label{tab:1}Example dynamics of semantic networks (increased $\uparrow$
or decreased $\downarrow$ after addition of a word).}

\begin{tabular}{|>{\raggedright}p{21mm}|>{\raggedright}p{38mm}|>{\centering}p{20mm}|>{\centering}p{20mm}|>{\centering}p{20mm}|>{\centering}p{20mm}|}
\hline 
\multirow{2}{21mm}{Initial set of 5~words} & \multirow{2}{35mm}{+ Additional word} & \multicolumn{4}{c|}{Semantic measures}\tabularnewline
\cline{3-6} 
 &  & Abstraction & Polysemy & Information content & Semantic similarity\tabularnewline
\hline 
\multirow{8}{21mm}{business, design, process, purpose, view} & + \{\} & 0.767 & 7.000 & 0.498 & 0.521\tabularnewline
\cline{2-6} 
 & + order & 0.769$\uparrow$ & 8.333$\uparrow$ & 0.477$\downarrow$ & 0.447$\downarrow$\tabularnewline
\cline{2-6} 
 & + competition & 0.769$\uparrow$ & 6.500$\downarrow$ & 0.500$\uparrow$ & 0.403$\downarrow$\tabularnewline
\cline{2-6} 
 & + time & 0.778$\uparrow$ & 7.500$\uparrow$ & 0.494$\downarrow$ & 0.408$\downarrow$\tabularnewline
\cline{2-6} 
 & + commercialization & 0.731$\downarrow$ & 6.000$\downarrow$ & 0.563$\uparrow$ & 0.430$\downarrow$\tabularnewline
\cline{2-6} 
 & + point & 0.778$\uparrow$ & 10.167$\uparrow$ & 0.477$\downarrow$ & 0.525$\uparrow$\tabularnewline
\cline{2-6} 
 & + issue & 0.769$\uparrow$ & 7.667$\uparrow$ & 0.505$\uparrow$ & 0.464$\downarrow$\tabularnewline
\cline{2-6} 
 & + dominance & 0.759$\downarrow$ & 6.500$\downarrow$ & 0.521$\uparrow$ & 0.417$\downarrow$\tabularnewline
\hline 
\end{tabular}
\end{table}

%3.5
\subsection{Main comparisons of individual design processes}

\begin{figure}[t!]
\begin{centering}
\includegraphics[width=\textwidth]{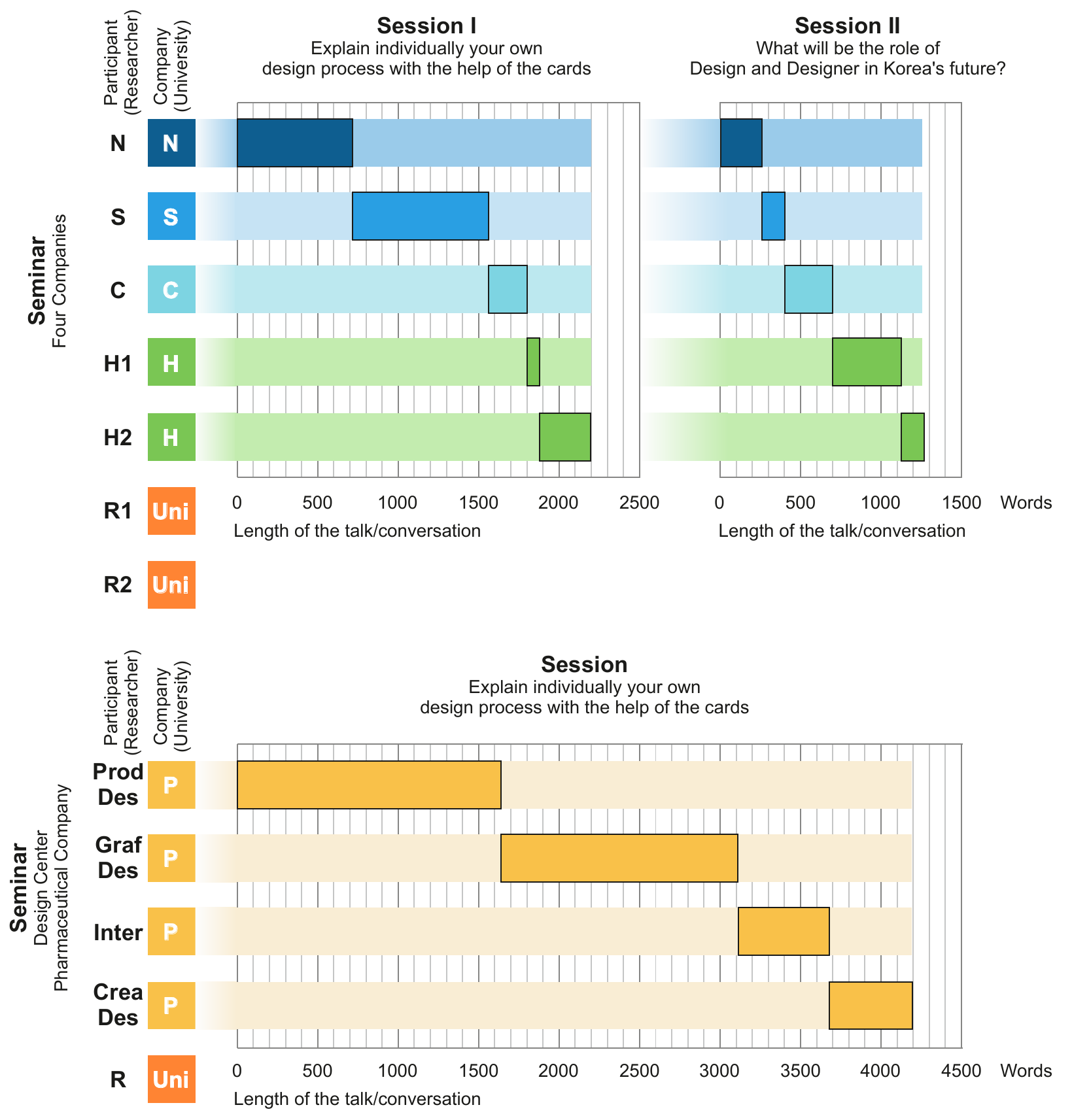}
\par\end{centering}
\caption{\label{fig:5}Sessions, their questions, sequence of talks by different
participants, and length of the whole talks per main speaker in words.}
\end{figure}

\begin{figure}[t!]
\begin{centering}
\includegraphics[width=\textwidth]{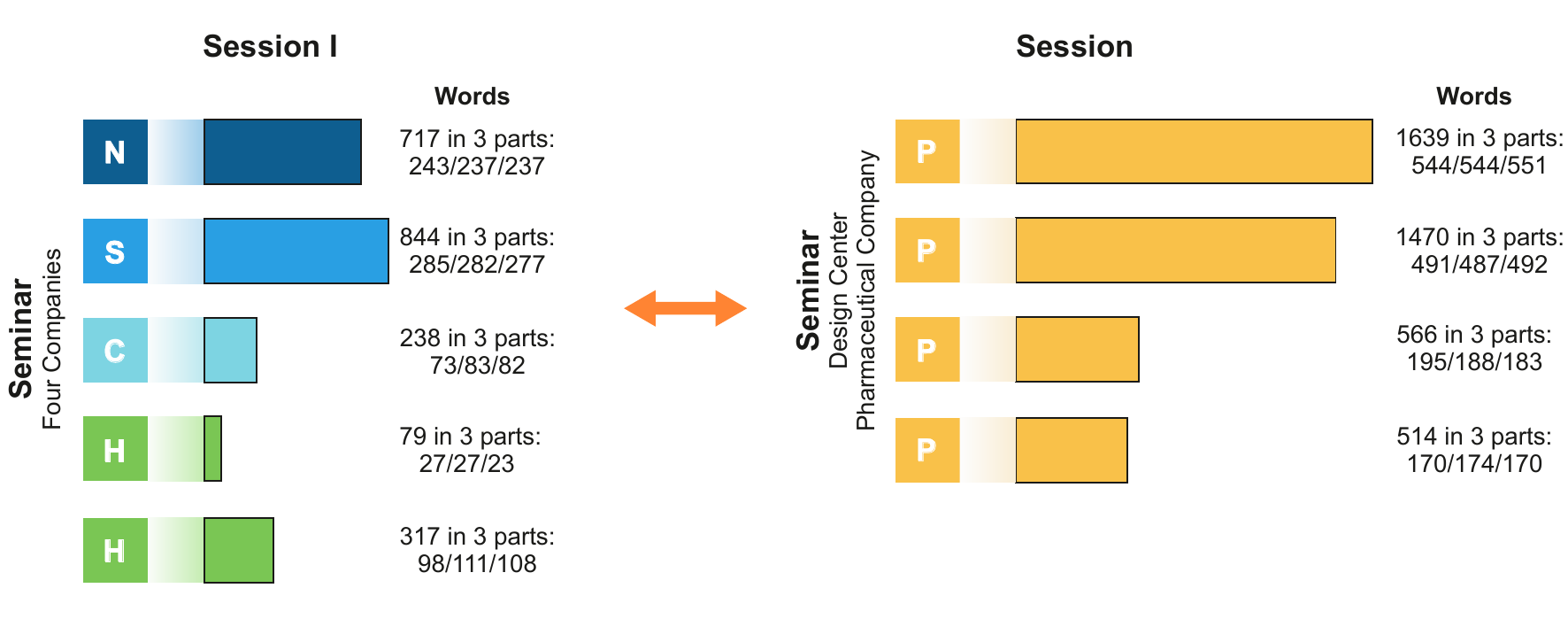}
\par\end{centering}
\caption{\label{fig:6}Sessions included in the analysis and division in three
temporal parts (in words). Shorter talks toward the end of the seminar appear to be due to fixed time duration of the design seminar and the lack of an external moderator that ensures equal presentation time for all participants.}
\end{figure}

We analyzed the sessions of the two seminars, focusing on individual explanations of their own design processes (Fig.~\ref{fig:5}). For the seminar with four companies, this was Session~I, whereas for the seminar with the design center of the pharmaceutical company, this was the sole session of the seminar (Fig.~\ref{fig:6}). As these explanations were not static and, in most cases, the main speaker received questions and comments, we explored the dynamics of these conversations.

A comparison of design thinking based on \emph{designer roles} was performed by dividing participants into two seminar groups: in the seminar with four companies, the participants were design managers, whereas in the seminar with the design center of the pharmaceutical company, the participants were specialized designers.

A comparison of design thinking based on \emph{company size} was performed by dividing participants into two other groups: participants N, C, H1 and H2 represent designers from small and medium-sized companies, whereas participants from the pharmaceutical company P1--P4 and company S represent designers from large companies. The criterion for a large company was an annual revenue exceeding USD~1~billion.

%3.6
\subsection{Statistics}

Statistical analyses of the constructed semantic networks were performed
using SPSS version~23 (IBM Corp., Armonk, NY, USA). The dynamics
of the semantic measures were evaluated using a two-factor ANOVA, where company
type was the first factor and time was the second factor. The Holm--Bonferroni
method was applied to correct for multiple comparisons \citep{Holm1979,Bonferroni1936}.
Both uncorrected and corrected $p$-values are reported in the main text.

%4
\section{Results}

%4.1
\subsection{Dynamics of abstraction in the individual design-thinking process}

\begin{figure}[t!]
\begin{centering}
\includegraphics[width=\textwidth]{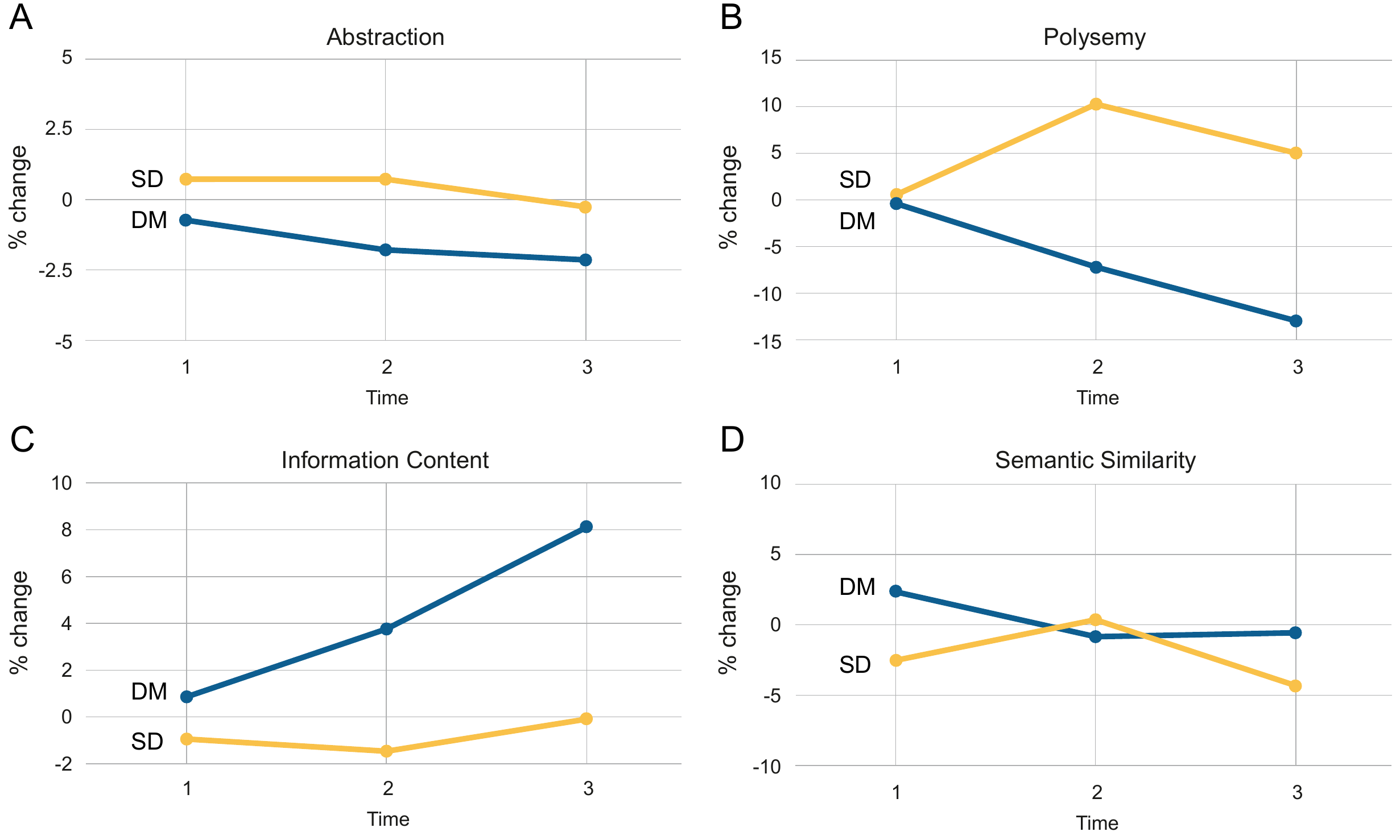}
\par\end{centering}
\caption{\label{fig:7}Dynamics of semantic measures in the individual design-thinking processes of specialized designers and design managers at the two seminars. Percent changes were plotted for abstraction (A), polysemy (B), information content (C) and semantic similarity (D) at 3 time periods corresponding to first, second or final third of the design conversations. DM, design managers in the seminar with four companies; SD, specialized designers in the seminar with design center of a company.}
\end{figure}

A comparison of the temporal dynamics of abstraction in the individual design-thinking processes of \emph{design managers} and \emph{specialized designers} showed statistically significant difference (two-factor ANOVA: $F_{1,21}=7.06$, uncorrected $p=0.015$; corrected $p=0.045$; Fig.~\ref{fig:7}A). The level of abstraction in the design thinking of the design managers in the seminar with four companies was lower at each time period; therefore, their attention appeared to focus on more concrete design issues.
The individual design-thinking processes of specialized designers in the seminar with the design center followed the same dynamic trend from higher initial abstraction to lower final abstraction; however, the abstraction was higher on average.
Notably, for both seminars, the focus on more concrete issues appeared toward the end of the design process. This temporal tendency toward decreased abstraction is consistent with a fuzzy front end in the initial stages of the design process and establishes a general pattern of addressing more abstract design aspects before proceeding to more specific ones.

A comparison of design thinking based on the company size showed similar dynamics of abstraction from high to low in both \emph{large} and \emph{small} companies; however, the level of abstraction was significantly higher for large companies at each time period, consistent with the employment of specialized designers (two-factor ANOVA: $F_{1,21}=9.76$, uncorrected $p=0.005$; corrected $p=0.02$).

%4.2
\subsection{Dynamics of polysemy in the individual design-thinking process}

The dynamics of polysemy in the individual design-thinking process of \emph{design managers} at the seminar with four companies and \emph{specialized designers} in the seminar with design center also exhibited statistically significant difference (two-factor ANOVA: $F_{1,21}=9.861$, uncorrected $p=0.005$; corrected $p=0.02$; Fig.~\ref{fig:7}B). The design thinking of design managers showed a temporal tendency toward decreased polysemy (an overall drop of $12.5\%$ at time period~3 compared to time period~1); therefore, the conversations appeared to use words with fewer meanings and lower ambiguity at the end. The trend of design thinking by specialized designers exhibits different characteristics, with an overall increase in polysemy in time periods~2 and 3 compared to period~1.
If we consider polysemy as an indication of a creative combination of ideas, one plausible interpretation is that the specialization of designers leads to a more creative combination, whereas designers with managerial roles are expected to solve concrete organizational problems that are more tightly constrained by available physical resources and realities.

A comparison of design thinking based on the company size showed that polysemy was significantly higher for large companies, which is consistent with the employment of specialized designers (two-factor ANOVA: $F_{1,21}=8.64$, uncorrected $p=0.008$; corrected $p=0.024$).

%4.3
\subsection{Dynamics of information content in the individual design-thinking process}

The individual design-thinking processes of \emph{design managers} in the seminar with four companies and \emph{specialized designers} in the seminar with the design center exhibited significant differences in information
content dynamics (two-factor ANOVA: $F_{1,21}=6.731$, uncorrected $p=0.017$; corrected $p=0.034$; Fig.~\ref{fig:7}C).
The temporal increase in information content in the design thinking appears to convey a higher amount of information regarding design issues and final design products.
The information content in the design thinking of design managers in the seminar with four companies increased by $7.2\%$ from time period~1 to time period~3. This increase in information content was substantially greater than the corresponding increase of $0.8\%$ observed for specialized designers.

A comparison of design thinking based on the company size showed that the information content increased over time in both groups; however, there was no statistically significant difference between the design-thinking processes of designers employed by \emph{large} or \emph{small} companies (two-factor ANOVA: $F_{1,21}=2.34$, $p=0.14$).

%4.4
\subsection{Dynamics of semantic similarity in the individual design-thinking process}

A comparison of the dynamics of semantic similarity in the design thinking of \emph{design managers} in the seminar with four companies and \emph{specialized designers} in the seminar with the design center showed no significant difference (two-factor ANOVA: $F_{1,21}=0.121$, $p=0.732$; Fig.~\ref{fig:7}D).
It should be noted, however, that in both seminars, semantic similarity decreased on average by 2.4~\% (with both groups combined) when estimated as the difference between the first and last time periods, thereby lending support to the occurrence of divergent thinking in the act of design creativity.

Similarly, the individual design-thinking processes were not significantly different based on the company size (two-factor ANOVA: $F_{1,21}=1.04$, $p=0.319$), lending support to the conclusion that design creativity does not depend on designer employment by \emph{large} or \emph{small} companies.

%5
\section{Discussion}

%5.1
\subsection{Abstraction, information content, and divergence in design thinking}

Changes in the four semantic measures based on WordNet~3.1 provide precise quantification of the temporal dynamics of the reconstructed individual design-thinking processes. The example provided in Table~\ref{tab:1} is derived from the beginning of the actual talk of the participant from company denoted by S in the seminar with four companies (Fig.~\ref{fig:5}). In this example, the introduction of the sixth word `time' in addition to the set of five initial words `business, design, process, purpose, view' introduces quantifiable changes in the four observed semantic measures, in particular, increase in abstraction and polysemy and decrease in information content and semantic similarity.
Such quantification appears to be in line with human judgments about abstraction, polysemy (multiplicity of meanings), information content carried by the addition of a new concept, and the similarity of the addition to the set of initial concepts.

In essence, with semantic measures, we have a practical tool that effectively quantifies characteristics that are, to a large degree, implicit for participants in a particular creative decision-making or design-thinking process. These quantitative characteristics are the missing ingredients required for successful reverse engineering of human creativity and its implementation in future AI~co-creative partner programs \citep{Kim2023}, which would assist human designers in inputting the right amount of divergence to make the design product desirable or attractive to the end user. The celebrated psychological relationship between divergent thinking and creativity \citep{Guilford1957,Hudson1974,Runco2020} is expected to be quantitative, rather than qualitative. For example, while insufficient divergence may be perceived as a lack of imagination, excessive divergence is indistinguishable from designer incompetence. Therefore, the narrow range of divergence falling within a few percentage points observed in the individual design-thinking processes of human designers (Fig.~\ref{fig:7}) could enable future AI programs to modify initial design solutions in a manner that mimics human creativity.

In research on design problem solving in the educational academic context, divergence of semantic similarity, increased information content, and decreased polysemy were found to predict the success of generated ideas \citep{Georgiev2018}.
Furthermore, it was demonstrated \citep{Taura2012,Georgiev2018} that the semantic approach rationalizes the distinction between ideas, thereby providing insights into creative design-thinking and decision-making processes.
The findings of the current study are consistent with the general trends observed in creative problem solving in an educational academic context \citep{Georgiev2018}. The individual design-thinking processes of both design managers and specialized designers exhibited temporal dynamics toward increased information content and decreased abstraction and semantic similarity (Fig.~\ref{fig:7}). Specialized designers in the seminar with the design center, however, exhibited a distinct increase in polysemy, compared to the decreased polysemy  observed for design managers in the seminar with four companies.
The results show possible differences between discussions in these two seminars: (1) conversations about design thinking and creative problem solving at a more general level and (2) conversations about design problem solving focused on specific creative ideas \citep{Georgiev2018}.
This outcome is in line with the observed differences between describing designing and designing. In particular, there were significantly more function-related and fewer structure-related issues in conversations describing designing compared to conversations involved in designing \citep{Lee2018}. This finding aligns with the requirement of the practical integration of disciplines \citep{Chen2022} and perspectives \citep{Selamat2021} as being essential for design creativity.

%5.2
\subsection{Theoretical and technological implications}

The theoretical implications concern the application of the method of semantic networks to quantify the dynamics of information exchange during conversations in design thinking. Dynamic quantification employs four semantic factors known to represent essential processes in design thinking. The main advantages of the method of semantic networks is that it allows for:
(1) quantification of design-thinking processes and connection with characteristics of the outcomes, such as creativity or innovation potential;
(2) comparison of different design-thinking approaches toward selection of an optimal one; and
(3) reverse engineering of AI co-creative partner programs to assist in the early stages of the design-thinking process, focused on divergent modification of given initial design solutions.
The dynamic quantification of essential processes in design thinking can automatize assessment techniques, for example, the most complex cognitive processes, such as the utilization of mental models \citep{Casakin2015}.
Thus, dynamic semantic networks provide a promising tool for monitoring
design thinking and enabling computer-supported reinforcement of creativity.
The development of AI co-creative partner programs, based on the prediction of human design-thinking trends might help refocusing efforts on higher creativity or innovation potential. Our recommendation for developers of creative AI is to mimic the quantitative dynamics of semantic measures reported in the individual design-thinking process of human designers and implement percentage divergence as a constraint on how much the AI output is allowed to deviate from the initial input given by the human designer.

%5.3
\subsection{Practical and societal implications}

This study is intended to contribute to the limited knowledge on the dynamics of individual design-thinking processes in a company context (Fig.~\ref{fig:2}) and differences in how design thinking and creativity are discussed in practice \citep{Lee2018}. Consequently, this study is the first to investigate the dynamics of discussions on design-thinking constructs by measuring: (1) changes in semantic measures, and (2) differences in conversations on design thinking in a company context.

The practical implications concern the reported differences in information
exchange with respect to designer roles and company size. These,
for example, can be focused on (1) synchronizing different articulations
of concrete design-thinking issues discussed by different participants; and 
(2) developing interventions with the help of AI co-creative partner programs, which record, process, and react to conversations that occur in the company, based on particular semantic factors in the information exchanged. The performance of these AI co-creative partner programs can be customized to the size of the companies and to different roles of designers in the companies. Interventions can achieve a higher amount of information and less ambiguous content, thereby improving performance. The integration of different perspectives on creativity and critical thinking has been identified as essential in design \citep{Spuzic2016}. Techniques specific to the application domain---in this case, design thinking and creativity---are instrumental to that goal.

The WordNet-based dynamic semantic network method reliably extracts quantitative data from transcribed English texts. The analyzed text could be the work of several authors and may reflect the information exchange between individuals in a social context. This makes the method easily reusable for researchers in the humanities, as language provides raw data for the study of a wide range of cognitive processes in a variety of social settings. Investing efforts in translating English WordNet into other languages would also enable this method to be applied to cultural studies.

%5.4
\subsection{Limitations of the study}

Recordings from design seminars in commercial companies are difficult to obtain, and the conditions under which recordings are performed are beyond the control of individual researchers. Nevertheless, the organizers of DTRS12 were able to collect and provide valuable design seminar data from five Korean companies. This allowed us to implement dynamic semantic networks to assess the association between divergent thinking and design creativity in a company context. The main comparison in the DTRS12 dataset that could be subjected to meaningful statistical analysis is between specialized design roles and design managers or between designers in large and small companies.
Although several semantic measures, including abstraction, polysemy, and information content, exhibited statistically significant changes in our analyses, our understanding could undoubtedly benefit from further research to enhance the generalizability of the reported findings given the limited sample size of the DTRS12 dataset.
One way to expand our findings is to collect more data from design conversations during the creation of actual consumer products and include companies from other countries. This would help in the assessment of possible cultural differences in design thinking \citep{Gong2023} and help determine whether perceived hierarchies (cf. Hofstede's power distance \citep{Hofstede1985}) in the workshops may have influenced the reports of individual design processes.

%6
\section{Conclusions}

The temporal dynamics of the semantic measures of abstraction, polysemy, and information content afforded us insight into how design and creativity issues are understood and possibly evolve in the course of design thinking and discussions about design.
From the constructed semantic networks of the two seminars, we quantified changes in abstraction, polysemy, information content, and overall semantic similarity in the conversations.
The results indicated that the individual design-thinking processes of design managers in the seminar with four companies exhibited significant differences in the dynamics of abstraction, polysemy, and information content compared to specialized designers in the seminar with the design center of single company.
The decrease in polysemy and abstraction and the increase in information content of the conversations about design in the seminar with the four companies demonstrates that the design thinking of design managers is focused on more concrete design issues, with a higher amount of information and less ambiguous content at the end.
Company size was not a factor that affected the amount of divergent thinking, which was quantified through a decrease in semantic similarity over time, lending support to the conclusion that designers in both large and small companies are equally creative.

\section*{Declaration of competing interest}

The authors declare that they have no known competing financial interests or personal relationships that could have appeared to influence the work reported in this paper.

\section*{Author contributions}

GVG: Conceptualization; Data curation; Funding acquisition; Investigation;
Methodology; Project administration; Resources; Software; Supervision;
Validation; Visualization; Writing - original draft. DDG: Conceptualization;
Data curation; Formal analysis; Investigation; Methodology; Resources;
Software; Validation; Writing - review \& editing.

\section*{Acknowledgments}

This research was partially funded by Academy of Finland 6Genesis Flagship grant number 346208, and by the Erasmus+ project ``Bridging the creativity gap'' (agreement number 2020-1-UK-01-KA202-079124).
We sincerely thank the editor and anonymous reviewers for taking the time to review our manuscript and providing constructive feedback to improve our presentation.

\section*{Data availability}

The authors do not have permission to share the DTRS12 dataset.



%%       \citet{<label>} ==> Jones et al. [21]
%%       \citep{<label>} ==> [21]
%%


\interlinepenalty=10000

\end{document}